# Multiconfigurational Nature of 5f Orbitals in Uranium and Plutonium Intermetallics

C. H. Booth[1], Yu Jiang[1], D. L. Wang[2], J. N. Mitchell[3], P. H. Tobash[3], E. D. Bauer[4], M. A. Wall[5], P. G. Allen[5], D. Sokaras[6], D. Nordlund[6], T.-C. Weng[6], M. A. Torrez[4], and J. L. Sarrao[7]

[1]Chemical Sciences Division, Lawrence Berkeley National Laboratory, Berkeley, California 94720, USA

[2]Nuclear Sciences Division, Lawrence Berkeley National Laboratory, Berkeley, California 94720, USA

[3]Materials Science and Technology Division, Los Alamos National Laboratory, Los Alamos, New Mexico 87545, USA

[4]Materials Physics and Applications Division, Los Alamos National Laboratory, Los Alamos, New Mexico 87545, USA

[5]Condensed Matter and Materials Division, Livermore National Laboratory, Livermore, California 94550, USA

[6]Stanford Synchrotron Radiation Lightsource, SLAC National Accelerator Laboratory, Menlo Park, CA 94025, USA

[7]Science Program Office- Office of Science, Los Alamos National Laboratory, Los Alamos, New Mexico 87545, USA

*Corresponding author*:
Corwin H. Booth
Chemical Science Division
MS 70A-1150
Berkeley, CA 94720 USA
Voice: 510-486-6079
FAX: 510-486-5596
Email: chbooth@lbl.gov
Web: http://lise.lbl.gov/chbooth

Draft, version 2.2, Last saved 5/25/2012, 3:14:15 PM, to be submitted to *PNAS*

Abstract:

**Uranium and plutonium's 5f electrons are tenuously poised between strongly bonding with ligand spd-states and residing close to the nucleus. The unusual properties of these elements and their compounds (eg. the six different allotropes of elemental plutonium) are widely believed to depend on the related attributes of f-orbital occupancy and delocalization, for which a quantitative measure is lacking. By employing resonant x-ray emission spectroscopy (RXES) and x-ray absorption near-edge structure (XANES) spectroscopy and making comparisons to specific heat measurements, we demonstrate the presence of multiconfigurational f-orbital states in the actinide elements U and Pu, and in a wide range of uranium and plutonium intermetallic compounds. These results provide a robust experimental basis for a new framework for understanding the strongly-correlated behavior of actinide materials.**

\body

The magnetic and electronic properties of actinide (An) materials have long defied understanding, where scientists prior to World War II (and even Mendeleev) placed the actinide series underneath the 5d transition series in the periodic table. The reason for such confusion is that the 5f orbital is intermediate between localized, as generally are the 4f orbitals in the lanthanide series, and delocalized, such as occurs in the d-orbitals of the transition metals. In those two limiting cases, well defined methodologies exist that account for their magnetic behavior, such as Hund's Rules, crystal-field theory, and quenched angular momentum theory. No similarly successful theory exists for the intermediate localization that occurs in elemental U, Np, and Pu, and their intermetallic compounds, and yet the consequences are likely fundamental toward understanding their complex behavior (1, 2).

Although the degree of f-electron localization is widely recognized as the dominant factor in determining the structural, magnetic, and electronic properties of the actinides, for instance, in determining basic crystal bonding (3), experimental methods for determining the f-orbital occupancy have generally failed to yield quantitative measurements, although some exceptions exist. For example, elemental Pu is thought to have an f-orbital occupancy near 5, close to that expected for a $5f^5$ ground-state configuration ($Pu^{3+}$), based on photoemission, $N_{4,5}$-edge x-ray absorption, and electron-energy loss spectroscopy (4, 5). In addition to these examples, several researchers have shown that so-called "two-fluid", or "dual nature", models of the 5f orbitals, whereby some fraction of the f-electrons contribute to delocalized behavior and the rest contribute to the more localized, local moment behavior, can successfully describe some properties (6), such as coexistent antiferromagnetism and superconductivity in $CeRhIn_5$ (7), the inelastic neutron scattering of $UPd_2Al_3$ (8), de Haas-van Alphen frequencies of $UPt_3$ (9), and the photoemission spectra of $PuCoGa_5$ and $PuIn_3$ (10). Such mixed behavior also manifests itself in the spin and orbital components of the angular momentum (11) —important quantities for understanding the absence of magnetism in plutonium (12). These results rely on fractional f-occupancies, especially in the delocalized channel. Recent Dynamical

Mean-Field Theory (DMFT) calculations by Shim, Haule, and Kotliar (13) suggest that, while the average f-occupancy is an important quantity, the actual ground state in elemental plutonium may require a more complete description. In particular, they find that, unlike cerium and ytterbium intermetallics which are described as dominated by two valence configurations ($f^0$ and $f^1$ for $Ce^{4+}$ and $Ce^{3+}$, $f^{13}$ and $f^{14}$ for $Yb^{3+}$ and $Yb^{2+}$), a description of elemental plutonium actually requires three valence configurations, $f^4$, $f^5$, and $f^6$. Here, we present both x-ray absorption near-edge structure (XANES) and resonant x-ray emission spectroscopy (RXES) data collected at the actinide $L_3$ edge in a wide variety of uranium and plutonium intermetallics that not only points to the necessity of a multiconfigurational ground state for understanding of elemental plutonium, but also demonstrates the wide applicability of such multiconfigurational ground states in actinide intermetallics in general.

There are some advantages to using $L_3$-edge spectroscopy for obtaining *f*-occupancies, especially for multiconfigurational states. At this X-ray absorption edge, a $2p_{3/2}$ core electron is excited into, primarily, a state of d symmetry, where the number of unoccupied 6d states is only a weak function of the f-occupancy (Fig. 1*a*). If a multiconfigurational f-state exists, its otherwise-degenerate components are split by the core-hole interaction, as the different number of f-electrons in each configuration screen the core hole differently. Since the total number of unoccupied d-states is approximately fixed, the excitation amplitude into any split states is proportional to that particular configuration's electronic occupancy. By associating a given peak with a particular configuration, the relative weight to each configuration can be simply determined. For instance, in RXES results on Yb intermetallics (14, 15), the integrated intensity $I_{13}$ and $I_{14}$ of features identified as due to the $f^{13}$ and $f^{14}$ configurations give the f-hole occupancy $n_f$ = $I_{13}/(I_{13}+I_{14})$. Similar methods have long been applied in XANES spectroscopy (16). While these measurements give the f-occupancy and configuration fractions in the excited state, which includes the core-hole and the outgoing photoelectron, such final-state occupancies are within several percent of those obtained using more sophisticated treatments for deep core-level excitations (17).

In this study, the An $L_3$-edge XANES data collected from α-U, α-Pu, and δ-Pu(1.9 at% Ga) along with 17 other uranium and 9 plutonium intermetallic samples delineate the correspondence between the edge position and localization of the 5f electrons. Pu $L_3$-edge XANES data are shown for typical Pu materials from this study in Fig. 1*b*. Similar U data are in Supporting Information (SI). The position of the main peak, known as the "white line" position, is shown in Fig. 1*c* and 1*d* as a function of the shift from the white-line position of the α-phase of the actinide (i.e. α-U or α-Pu), $\Delta E_\alpha$. Large shifts in $\Delta E_\alpha$ are observed, as are broadened white-line features for some compounds. Individual peaks are not observed, and so obtaining state configuration fractions is not possible from these data. The Sommerfeld coefficient to the linear component of the low-temperature specific heat, γ, is used as a measure of the degree of localization. The value of γ often is the defining quantity for heavy-fermion behavior, since it is proportional the effective carrier mass; that is, it is proportional to the density of states, $\wp$, at the Fermi level. In this sense, the flatter bands have a large linear specific heat, and are considered to have more localized character due to a higher f-orbital occupancy (18). This higher occupancy could be due to f-orbital hybridization with the conduction band, as in a Kondo model, or due to direct involvement of the f-band at the Fermi level, which is uncommon in the lanthanides. In the case of samples with magnetic transitions (many of the more localized materials are antiferromagnetic in their ground state), γ is determined at temperatures above any transitions to remove the effect of changes in magnetic degrees of freedom (19). Some of these transition temperatures are as high as 30 K, and so large errors are reported due to the larger contribution of phonon vibrations to the specific heat at such temperatures. (More information regarding the specific heat, including all the values of γ and transition temperatures, is available in SI.) As shown in Figs. 1*c* and 1*d*, there is a strong correlation between $\Delta E_\alpha$ and the degree of localization of the 5*f* electrons, as measured by γ. This correspondence is explained by considering that the higher f-occupancy (i.e., larger γ) implies more localized f-electrons are available for screening the $2p_{3/2}$ core hole, generating a more negative $\Delta E_\alpha$, as observed.

These XANES results indicate that the final-state shifts of the white line correlate well with a ground-state measurement of the density of states. While individual peaks are not observed in the white lines (Fig. 1*b*), there appears to be a correspondence between the width of the white line and the overall energy shift, consistent with two or more configurations, although possibly also indicating a broader 6d band.. Focusing on the U intermetallics, $\Delta E_\alpha > 6.5$ eV between the end-point samples, $UCd_{11}$ and α-U. Using energy shifts between known oxidation states, for instance, between $U^{3+}$ and $U^{4+}$ oxides, a change of one electron corresponds to about 4 eV. A similar value is found between Pu oxides (20). A 6.5 eV shift implies a change in f-occupancy of nearly 1.5 electrons. While this is possible, we point out that a ~1.5 eV shift is observed between $UPd_3$, which has an $f^2$ configuration and is one of the few An intermetallic materials measured with a relatively well-known *f*-occupancy (21), and $UO_2$, which is also $f^2$. We contend that such a shift is due to greater screening of the core-hole in $UPd_3$ due to conduction electrons (22). Such conduction electron screening reduces the Coulombic attraction between the core hole and the photoelectron, and may also affect excitations into the lower unoccupied d-states just above the Fermi level. Although this effect will be roughly constant between metals, it makes determining f-occupancy from the measured edge shifts less reliable.

To gain quantitative information about the valence in elemental U and Pu and their compounds, U and Pu $L_3$-edge RXES data (23-26) were collected at the An $L_{\alpha 1}$ emission line ($3d_{5/2} \rightarrow 2p_{3/2}$ corresponding to an emission energy $E_E$ of about 14.2 keV for Pu and 13.6 keV for U, see final state in Fig. 1*a*). Fig. 2 shows the X-ray emission spectra (XES) and the RXES for $UCd_{11}$ and α-Pu at several incident energies $E_I$ as a function of the transfer energy, $E_T = E_I - E_E$. Data on $PuSb_2$, $UCoGa_5$, and δ–Pu are available in the SI. Since excitations into the continuum imply that states are always available above a threshold energy and $E_E$ has a constant distribution for these states, $E_T \propto E_I$ for fluorescence lines. Excitations into unoccupied states with discrete energy levels, on the other hand, have a distribution with a constant $E_T$ as a function of $E_I$. The upper panels are well below the fluorescence threshold, and show excitations that are at approximately fixed $E_T$ as $E_I$ is increased in the lower panels, although the amplitude of the three individual features varies with $E_I$. It is important to note that excitations into states below the fluorescence threshold are not, in fact, discrete (although that is how they are referred to below) in these materials, but have a finite bandwidth, and so $E_T$ is expected to vary as much as a few eV. The bottom panels (Figs. 2*b* and 2*d*) show the fluorescence line as a dominant feature with a $E_T$ that will increase linearly as $E_I$ increases further.

These RXES data clearly show changes in lineshape due to multiple excitation features as a function of incident energy—made possible by the improved resolution of this technique (27), which is set by the $3d_{5/2}$ orbital (about 4 eV) rather than the $2p_{3/2}$ (between 7 and 10 eV), and the ability to separate excitations into the continuum. To determine the individual contributions to that lineshape, we follow standard procedures set forth by Dallera et al. (14, 15), and find a Lorentzian lineshape for both the fluorescence and the discrete excitation contributions. In general, three excitations are required to fit most of the data. Although $E_T$ varies by 1 or 2 eV with $E_I$ for the lowest-$E_T$ excitation (the $f^3$ U $L_3$ and the $f^6$ for Pu $L_3$), these three excitations remain well separated by about 4 eV in $E_T$, consistent with a difference of one electron occupancy for each state. The relative weights of each configuration and the fluorescence peak are shown in Fig. 3 as a function of $E_I$. The total configuration fractions are then obtained by integrating these results over $E_I$ (Table 1). Absolute errors are estimated by altering the lineshape for the standard discrete excitation, and are about 10% (one such alteration is discussed in the SI). Relative errors between these measurements are about 2%. (Unfortunately, the $PuSb_2$ data do not allow for such a determination, since the bandwidth-related shifts in $E_T$ are too large, see SI). While these below-threshold excitations allow for a measurement of the state configuration fractions and the overall f-occupancy, the RXES data also allow for the determination of the fluorescence threshold energy shifts. These shifts (Fig. 3*c* and 3*f*) indicate differences in the total screening of the core hole and include the effect of differences in the conduction electron density.

These results have important implications for understanding the nature of the ground states for all the measured actinide materials in Fig. 1. In particular, Pu in the α- and δ-forms is best described with partially delocalized and strongly multiconfigurational *f*-orbitals, both in observed changes in the XANES and the broad features in the RXES, each as compared to data from more localized samples such as $UCd_{11}$ and $PuSb_2$. Indeed, qualitative agreement is obtained with DMFT calculations (13) for the configuration $f^4$, $f^5$, and $f^6$ fractions in δ-Pu, which indicate about 60% $f^5$ configuration, compared to about (38±10)% measured here (see Table caption for discussion of error estimates). In addition, DMFT predicts a difference in the total *f*-occupation $\Delta n_f$=0.2 between α- and δ-Pu, whereas the RXES results give $\Delta n_f$=0.12±0.02 (Table 1). Furthermore, within the estimated absolute errors, a multiconfigurational ground state occurs even for our most localized actinide sample, i.e., $UCd_{11}$, raising the fundamental question of whether any true $U^{3+}$ intermetallic compound actually exists. Our results not only provide an accurate measure of the f-occupancy in plutonium for the first time, they advance a new paradigm for understanding the light actinides based upon a 5f-electron multiconfigurational ground state that goes far beyond a "dual nature" scenario.

**Materials and Methods**

**Sample preparation.** Metallic δ- and α-phase samples were first melted and then high temperature annealed to remove any lattice defects and He gas that accumulated while aging at room temperature. Subsequently, 2.3 mm diameter discs were punch pressed, lapped, and polished, using a succession of finer grit lapping films ending in a 1 μm surface finish and a final thickness of 70 μm. The samples were then dip-coated with liquid Kapton and cured at 150 °C for 2 hours. This encasement greatly reduces the oxidation of the metallic Pu over time. The final curing at 150 °C also reverts any potential damage or phase due to the mechanical polishing process. All sample preparation/processing was done in an inert atmosphere glovebox for safety and for the minimizing of the continuous oxidative nature of these materials.

Single crystals of all Pu intermetallic compounds were grown by the molten metal flux growth technique (28), as were single crystals of $UCoGa_5$, $UM_2Zn_{20}$ ($M$=Fe, Co, Ru), $USn_3$, $UCd_{11}$, and $U_2Zn_{17}$. Polycrystalline samples of $UAuCu_4$, $UAu_3Ni_2$, $UCu_5$, $UPt_3$, $UNi_2Al_3$, $URu_2Si_2$, $UPd_2Al_3$, and $UPd_3$ were synthesized by arc-melting the elements on a water-cooled Cu hearth with a Zr getter under an ultra-high pressure (UHP) Ar atmosphere. In some cases, the arc-melted samples were annealed under vacuum to improve crystallinity.

**XANES.** Nearly all of the X-ray data were collected in fluorescence mode on single solid pieces of material. Exceptions are XANES data from $PuCoGa_5$, $PuGa_3$, and $PuAl_2$. Each of these samples was ground with a mortar and pestle and passed through a 30 μm sieve. The resulting powder was mixed with boron nitride or, in the case of $PuAl_2$ only, brushed onto clear adhesive tape. XANES data from these powder samples were collected in transmission mode. The samples were loaded into a LHe-flow cryostat and data were collected with the samples near 30 K, although no changes with temperature have been observed up to room temperature. All fluorescence data in Fig. 1 were measured on BL 10-2 or 11-2 at the Stanford Synchrotron Radiation Lightsource (SSRL) over a period of 10 years, both before and after the upgrade to that facility that took place in 2003. All data were collected using a double-crystal Si(220) monochromator, half-tuned to remove unwanted harmonic energies from the X-ray beam. The fluorescence data were collected using a multi-element solid-state Ge detector and were corrected for dead time, and were also corrected for self-absorption using the program FLUO (29). All monochromator energies were calibrated to the first inflection point of the $L_3$ edge absorption from either $UO_2$ at 17166.0 eV (30) or $PuO_2$ at 18062.3 eV (20).

**RXES.** RXES data were collected at room temperature using the a 7-crystal Johann-type x-ray emission spectrometer (25) at the wiggler beamline 6-2 that incorporates a $LN_2$ cooled double-crystal Si(311) monochromator, and Rh-coated collimating and focusing mirrors. U $L_{\alpha1}$ (13.6 keV) emission was measured using Ge(777) analyzer crystals, and Pu $L_{\alpha1}$ (14.2 keV) emission was measured using Si(777) analyzer crystals. The analyzer energy was calibrated using the nearby elastic peak from the 999 reflection and the already-calibrated monochromator energy. The resolution was measured from the elastically scattered beam to be 1.4 eV and 1.7 eV, respectively, at the An $L_{\alpha1}$ emission energies.

Determination of the lineshapes follows the methods of Dallera et al. (14). The intrinsic lifetime broadening (27) of any observed features is set by the $3d_{5/2}$ orbital (about 4 eV), rather than the $2p_{3/2}$ (between 7 and 10 eV). The fluorescence peak lineshape and position were determined at an $E_I$ that was well above threshold. The U $L_3$ edge data were fit using $E_{L\alpha 1}$ = 13616.1 eV ($UCd_{11}$) or 13617.1 eV ($UCoGa_5$), and a width $\Gamma_F$ = 5.9 eV. The Pu $L_3$ edge data were fit using $E_{L\alpha 1}$ = 1477.7 eV and a width $\Gamma_F$ = 6.5 eV. The normalized emission lineshape from the discrete excitations was obtained well below threshold for the most localized samples measured, namely $UCd_{11}$ and $PuSb_2$. These data were fit with a skewed Lorentzian:

$$\frac{I_E}{I_0} = W \frac{\Gamma_S^2}{\pi[(E_T - \langle E_T \rangle)^2 + \Gamma_S^2]} \{1 + erf\, [\frac{\alpha(E_T - \langle E_T \rangle)}{\sqrt{2}\Gamma_S}]\} \,,$$

where $W$ is the weight coefficient at fixed $\langle E_T \rangle$ for a given $E_I$ and $erf$ is the error function. The excitation width is $\Gamma_S$ = 3.3 eV and the skew parameter is $\alpha$=0.29 for the U edge data and $\Gamma_S$ = 4.7 eV and $\alpha$ = 0.30 for the Pu edge data.

Once the lineshape parameters were determined, an average $\langle E_T \rangle$ was found for each of the three discrete states as a function of $E_I$. These values were then fixed for the reported results at all $E_I$, except for the first peak (the $f^3$ peak for U edge and the $f^6$ peak for the Pu edge data). As noted in the text, this energy shifts by 1 to 2 eV, possibly due to a broad band for this configuration. This shift is severe enough in the $PuSb_2$ data so as to make it impossible to fit for individual f-configurations (see SI). Also note that the width of each peak is due to the convolution of the spectrometer and the final-state lifetime. Difference cuts along $E_I$ showed the broadening by the intermediate-state lifetime and the mono.

In addition, below threshold, the fluorescence peak was held to zero amplitude to avoid correlations with the other peaks, mostly the $f^4$ peak for the Pu $L_3$-edge RXES data. This constraint is not required if $E_T$ is not allowed to vary as a function of $E_I$, and such fits generate configuration fractions within the stated error estimates, although the fits are of poorer quality. However, the fluorescence threshold in these fixed-$E_T$ fits is less well defined than shown in the floating-$E_T$ fit results in Fig. 3c and 3f. An important improvement to these methods will be better defining the relationship between the fluorescence peaks and the below-threshold peaks.

As a final note, an acceptable (but much lower quality) fit can be obtained to the entire RXES spectra with the full Kramers-Heisenberg (KH) formula (23, 25) using three discrete and three fluorescence peaks, giving lower estimates of $n_f$ (eg. $n_f$=5.2 for δ-Pu and 5.0 for α-Pu). We attribute the low quality of these fits to the bandwidth effects in determining $E_T$ (not accounted for in such fits) and the use of perturbation theory in deriving the KH formula.

**Acknowledgements.** Work at Lawrence Berkeley National Laboratory supported by the Director, Office of Science, Office of Basic Energy Sciences (OBES), of the U.S. Department of Energy (DOE) under Contract No. DE-AC02-05CH11231. Work at Los

Alamos National Laboratory (LANL) was performed under the auspices of the U.S. DOE, OBES, Division of Materials Sciences and Engineering and funded in part by the LANL Directed Research and Development program. X-ray absorption and RXES data were collected at the Stanford Synchrotron Radiation Lightsource, a national user facility operated by Stanford University on behalf of the DOE, Office of Basic Energy Sciences. The authors acknowledge enlightening conversations with J. A. Bradley, G. Kotliar, G. H. Lander, J.-P. Rueff, J. Seidler, J. D. Thompson, and Z. Fisk.

**Figure legends**

**Fig. 1.** Actinide $L_3$-edge follows strongly correlated electron behavior. (*a*) Energy level diagram demonstrating the *dominant* transitions and the final state splitting due to the generation of the core hole, as well as the final states relationship to $E_I$ and $E_T$ for the RXES data. Note that other decay channels between the intermediate and final state configurations also occur (for instance, from a $f^4$ intermediate state to an $f^5$ final state); we have chosen to only illustrate the dominant channels for clarity. The intermediate states in the middle of the diagram represent the final state for the XANES data shown in (*b*). (*b*) Representative results from Pu $L_3$ edge XANES spectroscopy. (*c,d*) Linear coefficient of the low temperature specific heat in the normal state, γ, as a function of the shift of the peak in the white line relative to the α-An sample. Many of the γ values come from the literature (12, 19, 31-48). These results are available in tabular form in SI, together with individual references for the γ values.

**Fig. 2**. XES and RXES data on actinide intermetallics. Representative XES (*a*) and RXES (*b*) for $UCd_{11}$, as an example of a strongly localized An intermetallic. (*c,d*), Analogous data for α-Pu, as an example of a more delocalized An intermetallic. The colors in the RXES data represent the normalized emission flux. Note the clearly sharper resonance in both the XES and RXES (yellow peak) plots for the $UCd_{11}$ compared to the α–Pu data. Similar results for $UCoGa_5$, $PuSb_2$, and δ-Pu are available in SI.

**Fig. 3.** Multiconfigurational orbital weights. Relative weights of the principle components to (*a,b*) the discrete-state excitation and the (*c*) fluorescence spectra for the measured U intermetallic samples. (*d-f*), Analogous data from the Pu intermetallics. In the case of the U intermetallics, we assign the excitations at transfer energies of 3548 eV, 3552 eV, and 3556 eV to $f^3$, $f^2$, and $f^1$ configurations based on comparisons to the oxide. Likewise for the Pu intermetallics, the excitations at 3783 eV, 3787 eV, and 3791 eV are assigned to $f^6$, $f^5$, and $f^4$ configurations. The threshold energy shift in (*c*) is about 2.2 eV, and in (*d*) is about 4.5 eV. Discrete-state peak coefficients for $PuSb_2$ are somewhat different, and are available in SI.

**Table 1.** *f*-orbital occupancy and configuration fraction measurements

| Sample | $n_f$ | Configuration fractions | | |
|---|---|---|---|---|
| | | $f^1$ | $f^2$ | $f^3$ |
| UCd$_{11}$ | 2.71 | 0.07 | 0.15 | 0.78 |
| UCoGa$_5$ | 1.92 | 0.32 | 0.44 | 0.24 |
| | | $f^4$ | $f^5$ | $f^6$ |
| δ-Pu (1.9% Ga) | 5.28 | 0.17 | 0.38 | 0.45 |
| α-Pu | 5.16 | 0.19 | 0.46 | 0.35 |

Values of the f-orbital occupancies, $n_f$, are determined by the weighted sum of the integrated intensities from each configuration peak in Fig. 3*a*, *b*, *d*, and *e*, for example, $n_f = (4I_4 + 5I_5 + 6I_6)/(I_4 + I_5 + I_6)$, where $I_4$ is the integrated intensity of the $f^4$ peak, etc. Absolute errors are estimated by altering the lineshape for the standard discrete excitation, and are about 10%. Relative errors between these measurements are about 2%.

# Supplementary Information for "Multiconfigurational Nature of 5f Orbitals in Uranium and Plutonium Intermetallics"

C. H. Booth[1], Yu Jiang[1], D. L. Wang[2], J. N. Mitchell[3], P. H. Tobash[3], E. D. Bauer[4], M. A. Wall[5], P. G. Allen[5], D. Sokaras[6], D. Nordlund[6], T.-C. Weng[6], M. A. Torrez[4], and J. L. Sarrao[7]

[1]Chemical Sciences Division, Lawrence Berkeley National Laboratory, Berkeley, California 94720, USA

[2]Nuclear Sciences Division, Lawrence Berkeley National Laboratory, Berkeley, California 94720, USA

[3]Materials Science and Technology Division, Los Alamos National Laboratory, Los Alamos, New Mexico 87545, USA

[4]Materials Physics and Applications Division, Los Alamos National Laboratory, Los Alamos, New Mexico 87545, USA

[5]Condensed Matter and Materials Division, Livermore National Laboratory, Livermore, California 94550, USA

[6]Stanford Synchrotron Radiation Lightsource, SLAC National Accelerator Laboratory, Menlo Park, CA 94025, USA

[7]Science Program Office- Office of Science, Los Alamos National Laboratory, Los Alamos, New Mexico 87545, USA

**Summary of file contents**

These materials begin with the *Specific heat* section, which includes basic characterization details of the measured materials, focusing on the specific heat data (Fig. S1 and Table S1). The *Other Supporting Text and Figures* section shows some of the U $L_3$ XANES data in Fig. S2, and the available RXES and XES data from all the samples not shown in the main article (Figs. S3-S5). Finally, a RXES simulation generated from a model density of states is reported in *Example of Potential Absolute Errors and Lineshapes*, which is then fit using the skewed-Lorentzian method in the manuscript to demonstrate the method's efficacy for discerning multiconfigurational states (Figs. S6-S7).

**Specific heat and other characterization**

Most of the specific heat data reported in Table S1 and Fig. 1 has been previously published. An example of data collected for this work for PuSb$_2$ and PuIn$_3$ is shown in Fig. S1, together with the measured magnetic susceptibility. The specific heat data were collected on a Quantum Design Physical Property Measurement System (PPMS) and fit using the standard form $C/T = \gamma + \beta T^2$. Since we are interested in the Kondo contributions to $C/T$, we do not use data when the sample is in a magnetically ordered states, as the reduction in entropy adversely affects the relationship between charge localization (Kondo temperature) and $\gamma^1$. We therefore fit data above any magnetic transitions. The fits to data from PuSb$_2$ and PuIn$_3$ (Fig. S1) are examples of this situation. For literature data where only the low-temperature $\gamma$ is reported, we refit the data as just described. Other samples where magnetic transitions occur include Pu$_2$Ni$_3$Si$_5$ and PuGa$_3$. A source of error arises if the data above, say, $T = \Theta_D/10$, where $\Theta_D$ is the Debye temperature derived from $\beta$. By fitting $C/T$ data from samples that do not order magnetically with similar $\Theta_D$, such as YbAgCu$_4$, we find that fits to near 20 K are about 10% high compared to fits to the low-temperature data. For all such fits reported here, we attached an error of 100 mJ mol$^{-1}$ K$^{-2}$ as a conservative estimate. Note that all data reported here are not, where possible, corrected for the lattice contribution via subtraction using a nonmagnetic analogue, since much of the literature data would not be directly comparable.

**Other Supporting Text and Figures**

Representative U L$_3$ XANES data are shown in Fig. S2. RXES and XES data for δ-Pu are shown in Fig. S3. These data are very similar to the data from α-Pu (Fig. 2). Data on UCoGa$_5$ are shown in Fig. S4, and data from PuSb$_2$ are shown in Fig. S5. These latter data have some different properties, as discussed forthwith.

The data from PuSb$_2$ (Fig. S5) shows a single discrete excitation at energies below the fluorescence threshold. In contrast to data from the other measured materials, the energy position of this peak does not stay roughly constant with $E_I$ (Fig. S5d), but instead shifts from about 3778 eV to 3782 eV while the peak has significant weight in the spectrum (Fig. S5c). Over this energy range, the peak width does not change, and is, in fact, quite narrow as mentioned above, with $\Gamma_S = 4.7$ eV. The lack of change and overall sharpness of this peak does not allow for an interpretation with a significant multiconfigurational state; however, due to the large energy shift, we cannot uniquely assign the *f*-occupancy of this single configuration. The shift of the single-state $E_T$ requires further study, but is likely due to a larger bandwidth in the unoccupied *d*-states below the fluorescence threshold than in the other measured samples.

**Example of Potential Absolute Errors and Lineshapes**

The procedure used for fitting the data discussed in this work involves fitting skewed Lorentzians to the features in the XES data, as discussed in the main work. The absolute quoted estimated error of 10% was estimated by using different, generally less satisfactory lineshapes. Here we give an example of one particular method that eventually may lead to better lineshapes. In this example, we choose a single-configurational density of unoccupied electronic states (DOS) and then fit this simulated data using the methods used in the paper to fit the data.

The chosen DOS is shown in Fig. S6a, and is comprised essentially of a step function added to a Lorentzian. This DOS is then used to generate the PFY in Fig. S6b and a full RXES spectrum, as shown in Fig. S6c, by making use of the KH formula and following Eq. 19 in Rueff and Shukla's review article (21):

$$I_E \propto \sum \sum \int d\varepsilon \frac{\eta(\varepsilon)}{(E_{ig} + \varepsilon - E_I)^2 + \Gamma_i^2/4} \times \frac{\Gamma_f/2\pi}{(E_{fg} + \varepsilon - E_T)^2 + \Gamma_f^2/4},$$

where $\eta(\varepsilon)$ is the DOS and $\varepsilon$ is the energy with respect to the Fermi level, $\Gamma_f$ and $\Gamma_i$ are the width of the final and intermediate states, and $E_{fg}$ and $E_{ig}$ are the threshold energies. Interference terms are neglected here. These values were chosen to reflect data for the U $L_{III}$ edge.

As in the paper, we determine the so-called fluorescence lineshape fit parameters by fitting to these simulated data well above the edge, and the discrete lineshape parameters well below the edge. The inclusion of a sharp peak in the DOS was determined to be necessary to obtain simulated data well below the edge that is not overly skewed. The skew parameter $\alpha \approx 0.23$ in this simulation, close to that found for fits to all the real data ($\alpha \approx 0.30$). It is important to note that a step-function DOS without the sharp peak gave a much larger skew parameter ($\alpha \approx 9$).

One the lineshapes were determined, we fit the simulated XES data assuming three peaks: the fluorescence peak and two discrete peaks. The energy of the second peak is fixed at 4 eV above the main peak to avoid it floating uncontrollably in these simulation fits. Fit examples are shown in Fig. S7a and Fig. S7b. Fitting all the XES data at all EI and results in amplitudes for each peak as a function of EI shown in Fig. S7c. Indeed, adding the second discrete peak does improve the fit. However, the integrated area under the curves indicates the first discrete peak (P2) accounts for 91% of the total amplitude of P2 and P3. Since the DOS is, in fact, a single configuration, the fit results indicate that the method is consistent with a single configuration with 10%, consistent with the stated absolute error in the paper.

**Figure legends**

**Figure S1 | Specific heat examples**. **a,b,** Specific heat data from (a) PuSb$_2$ ($\gamma$ = 397 mJ mol$^{-1}$ K$^{-2}$) and (b) PuIn$_3$ ($\gamma$ = 326 mJ mol$^{-1}$ K$^{-2}$), as examples where a linear fit was extrapolated from above an antiferromagnetic transition at relatively high T.

**Figure S2 | Representative U L$_3$ edge XANES data for several samples.** All these data are collected at T ≈ 30 K.

**Figure S3 | RIXS results for $\delta$-Pu**. **a,b,** RIXS data (a) and XES at several incident energies (b) for $\delta$-Pu.

**Figure S4 | RIXS results for UCoGa$_5$. a,b,** RIXS data (a) and XES at several incident energies (b) for UCoGa$_5$.

**Figure S5 | RXES results on PuSb$_2$. a,** XES at several incident energies for PuSb$_2$. **b-d,** RXES data (b) and fit results showing the single discrete excitation (c) amplitude and (d) peak transfer energy.

**Figure S6 | Single-configuration simulation. a,** DOS and PFY of simulated RXES. The partial fluorescence yield (PFY) shows the diagonal cut through the RXES simulated data in (b). **b,** RXES simulation. **c,** XES cuts at fixed E$_I$.

**Figure S7 | XES and fit results for single-configuration simulation. a,** XES and fit results for E$_I$=17160 eV to simulation in Fig. S6. P1 indicates the fluorescence peak, and P2 and P3 are discrete excitations 4 eV apart. Fit results to this simulation follow the methods outlined in the main paper. **b,** XES and fit results for E$_I$=17170 eV of simulation in Fig. S6. **c,** Amplitudes for each fitted peak as a function of E$_I$. Integrated amplitudes indicate P2 makes up 91% of the amplitude of P2+P3.

**Table S1.** Data used to generate Fig. 1, together with the literature source for the specific heat results. All values of γ are obtained above any noted transitions. The ground state of each sample is also noted: sc=superconductor, pm=paramagnet, afm=antiferromagnet, sf=spin fluctuation, sg=spin glass, and qo=quadrupolar order. Specific heat data are per mol An. $L_3$ peak position errors are less than 0.1 eV, and are calibrated to the first inflection point of the dioxide at 17166.0 eV for $UO_2$ and 18062.3 eV for $PuO_2$.

| compound | ground state | $L_3$ peak position (eV) | $\gamma$ (mJ·mol$^{-1}$·K$^{-2}$) | $\gamma$ source |
|---|---|---|---|---|
| α-U | sc (0.5 K) | 17173.1 | 9.13 | (1) |
| $UCoGa_5$ | pm | 17172.8 | 21 | this work |
| $USn_3$ | pm | 17172.5 | 172 | (2) |
| $UAuCu_4$ | afm (30 K) | 17171.9 | 100 | (3) |
| $UAl_2$ | sf | 17171.9 | 70[a] | (4) |
| $URu_2Zn_{20}$ | pm | 17171.7 | 190 | (5) |
| $UAu_3Ni_2$ | sg (3.6 K) | 17171.6 | 270 | (3) |
| $UCo_2Zn_{20}$ | pm | 17171.6 | 151[e] | this work[e] |
| $UFe_2Zn_{20}$ | pm | 17171.5 | 176 | (6) |
| $UCu_5$ | afm (15 K) | 17171.5 | 203 | this work |
| $UPt_3$ | sf | 17171.4 | 225 | (7) |
| $UNi_2Al_3$ | afm (4.6 K) | 17171.4 | 129 | (8) |
| $UAuPt_4$ | sf | 17171.1 | 260 | (3) |
| $URu_2Si_2$ | afm (17.5 K) | 17171.0 | 180 | (9) |
| $U_2Zn_{17}$ | afm (9.7 K) | 17170.4 | 412 | (10) |
| $UPd_2Al_3$ | afm (14 K), sc (2 K) | 17170.4 | 210 | (11) |
| $UPd_3$ | qo (7.5 K) | 17170.4 | 280[d] | (12) |
| $UCd_{11}$ | afm (5 K) | 17166.5 | 840 | (13) |
| $UO_2$ | afm (30.8 K) | 17172.2 | - | |
| α-Pu | pm | 18068.2 | 17 | (14) |
| δ-Pu (1.9 at% Ga) | pm | 18066.9 | 42 | (15) |
| $PuCoGa_5$ | sc (18.2 K) | 18066.1 | 130[b] | (16) |
| $Pu_2PtGa_8$ | pm | 18065.3 | 51 | this work |
| $PuAl_2$ | afm (3.5 K) | 18065.2 | 260 | (17) |
| $PuCoIn_5$ | sc (2.7 K) | 18064.7 | 231 | this work |
| $PuPt_2In_7$ | pm | 18064.4 | 250 | this work |
| $PuGa_3$ | afm (24 K) | 18064.4 | 412[c,d] | (18) |
| $Pu_2Ni_3Si_5$ | afm (35 K) | 18064.3 | 476[c] | this work |
| $PuIn_3$ | afm (14.6 K) | 18064.2 | 325 | this work |
| $PuSb_2$ | afm (20 K) | 18063.5 | 397[c] | this work |
| $PuO_2$ | pm | 18068.4 | - | |

[a] γ determined at high magnetic field to reduce spin fluctuations.
[b] An alternate value is 77 mJ mol$^{-1}$ K$^{-2}$ (19). The change at $T_c$ gives 95 mJ mol$^{-1}$ K$^{-2}$ (20)
[c] These values for γ are extrapolated from data above 20 K due to relatively high Néel temperatures, and so are less reliable. These data are marked with 100 mJ mol$^{-1}$ K$^{-2}$ error bars in Fig. 1.
[d] These values for γ are obtained by refitting the data in the cited reference above the listed phase transition.
[e] Fits to $C/T$ data by Wang et al. (5) give γ = 558 mJ mol$^{-1}$ K$^{-2}$ at $T$=0 K, but this result does not hold for higher $T$. The presence of antiferromagnetic fluctuations likely enhances γ. The value reported here is for a higher temperature fit between 10-15 K on data from a new sample.

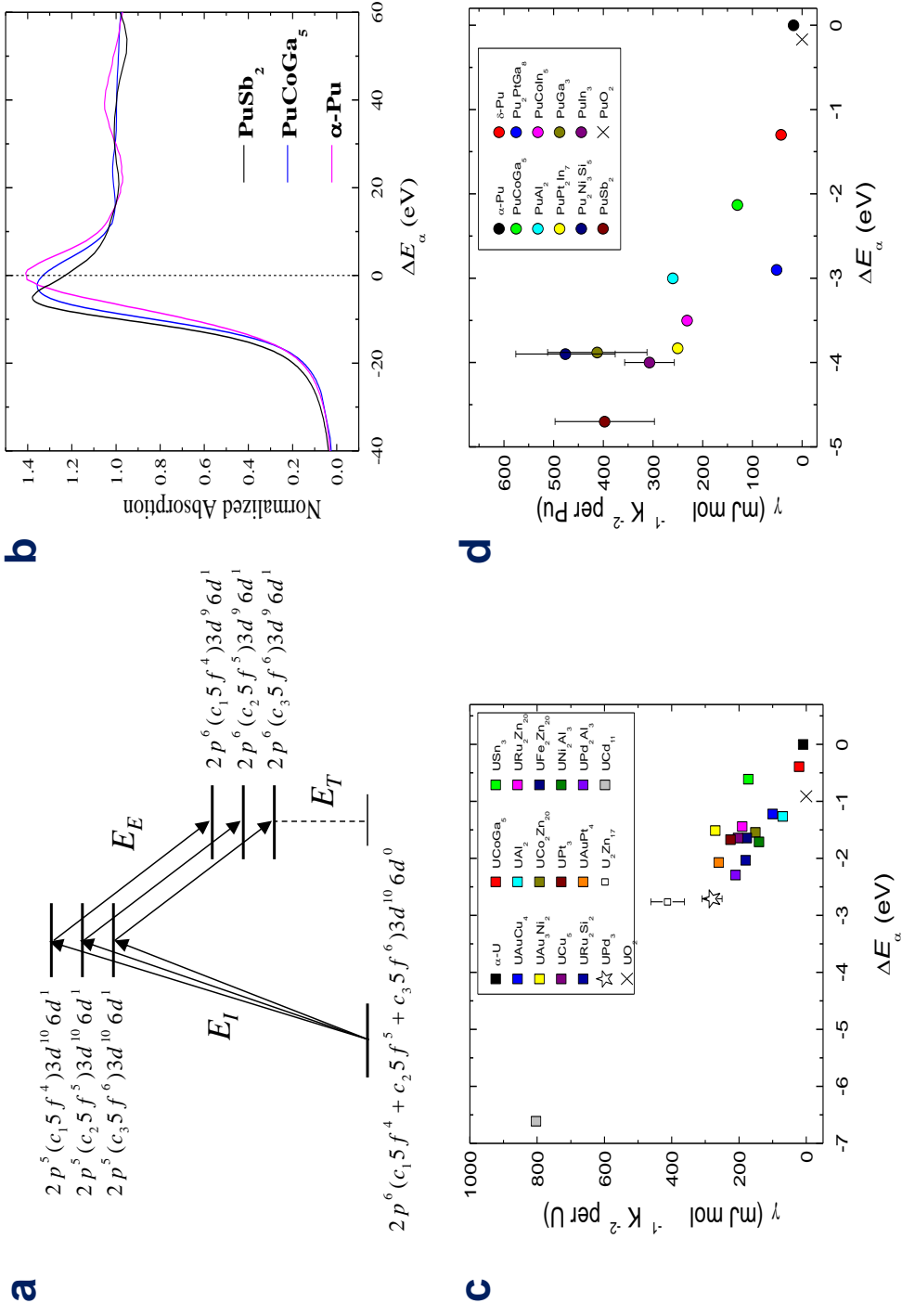

Figure 1

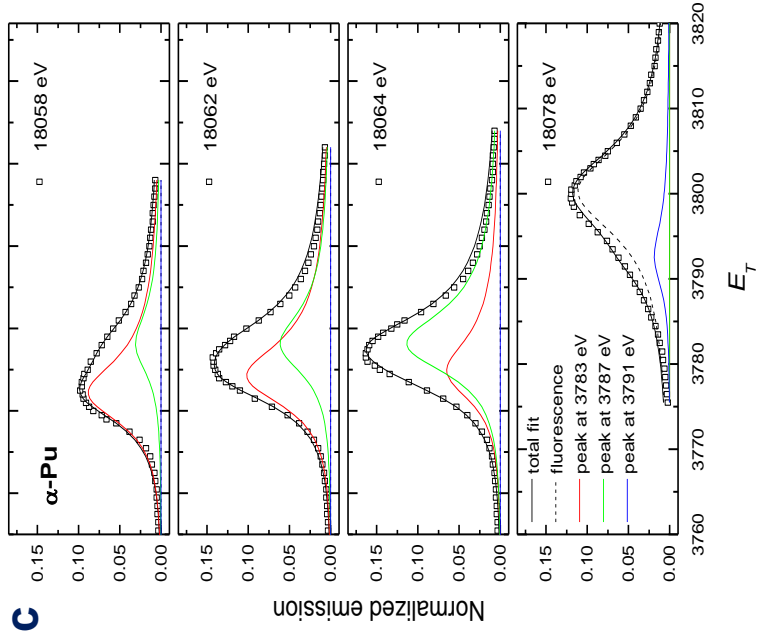
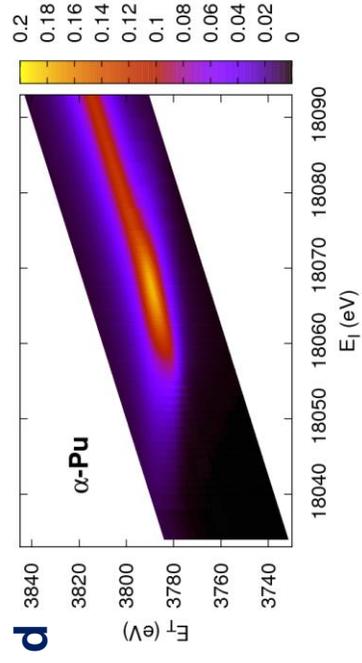
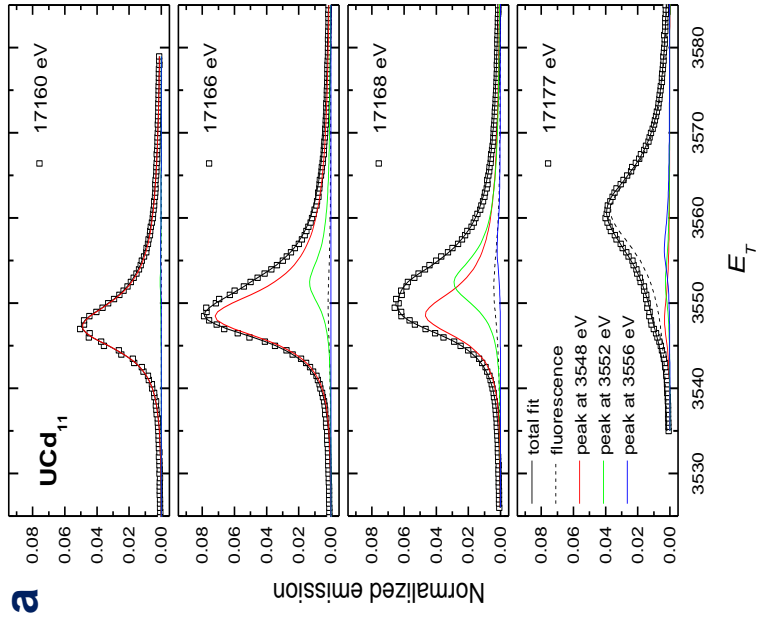
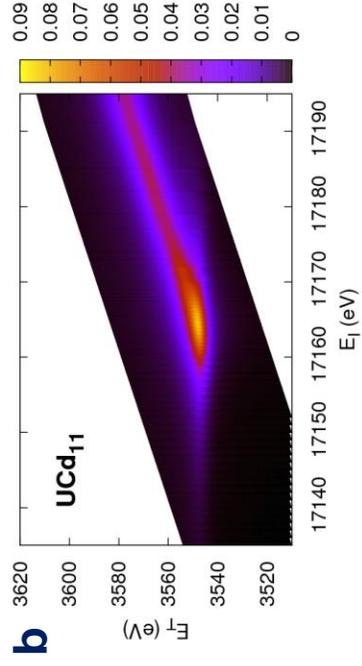

**Figure 2**

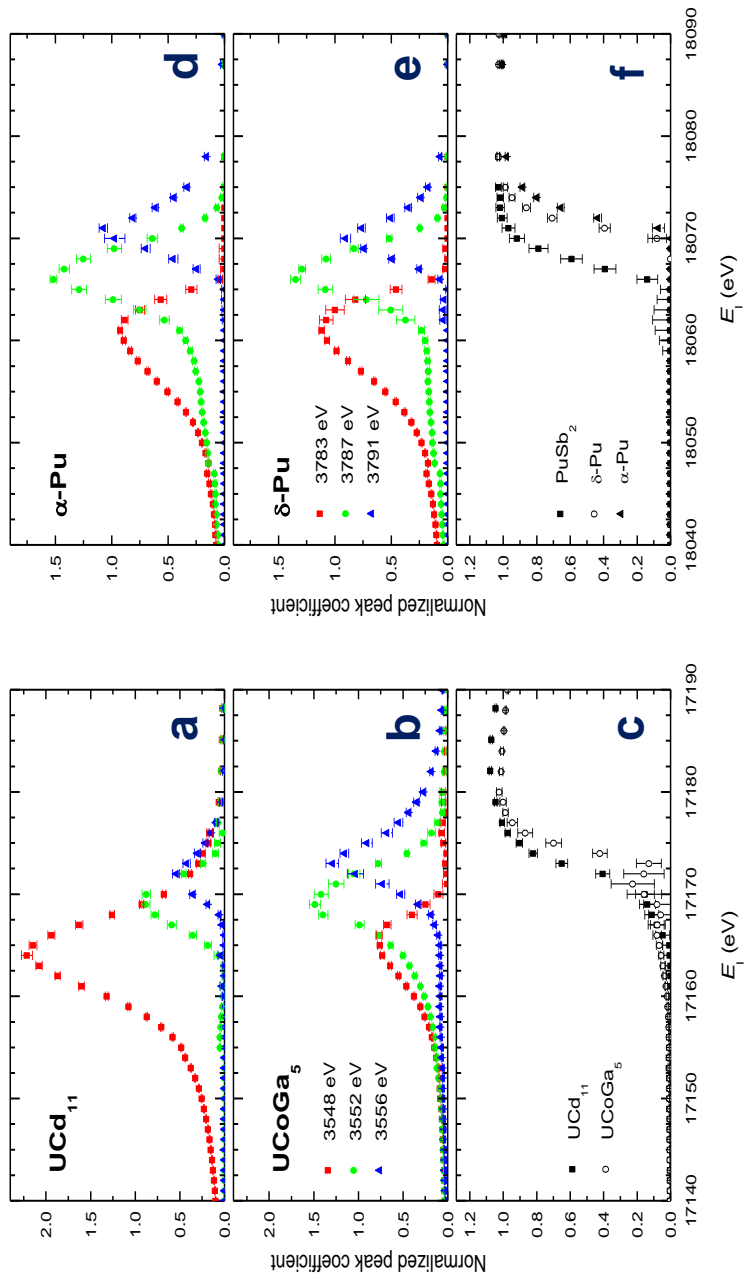

Figure 3

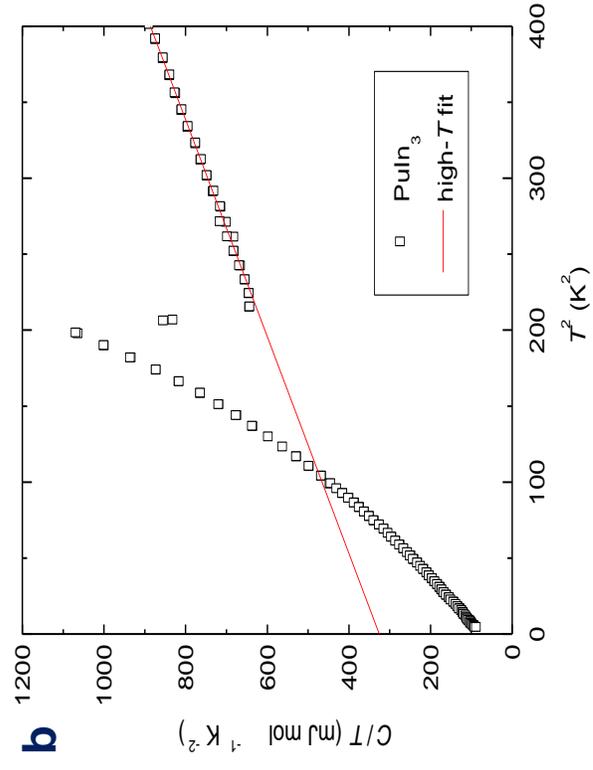
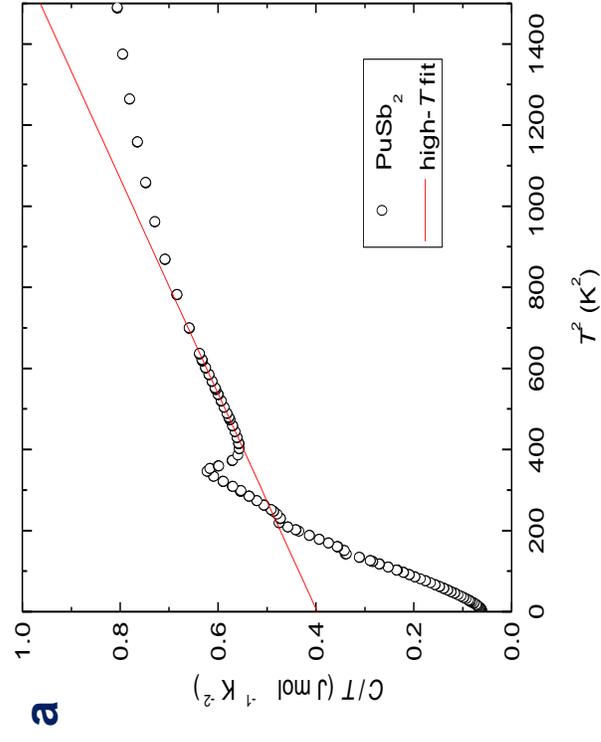

Figure S1

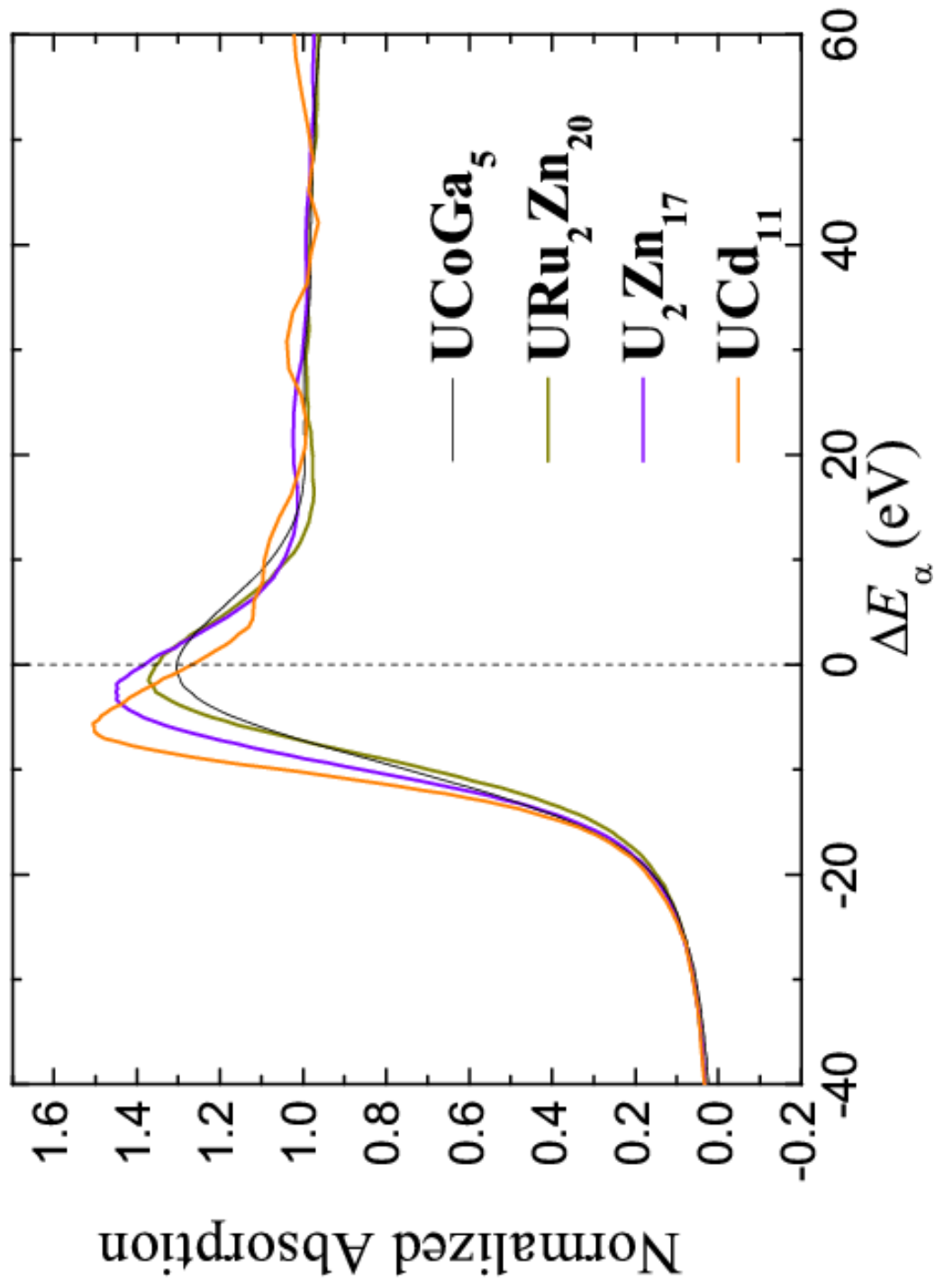

Figure S2

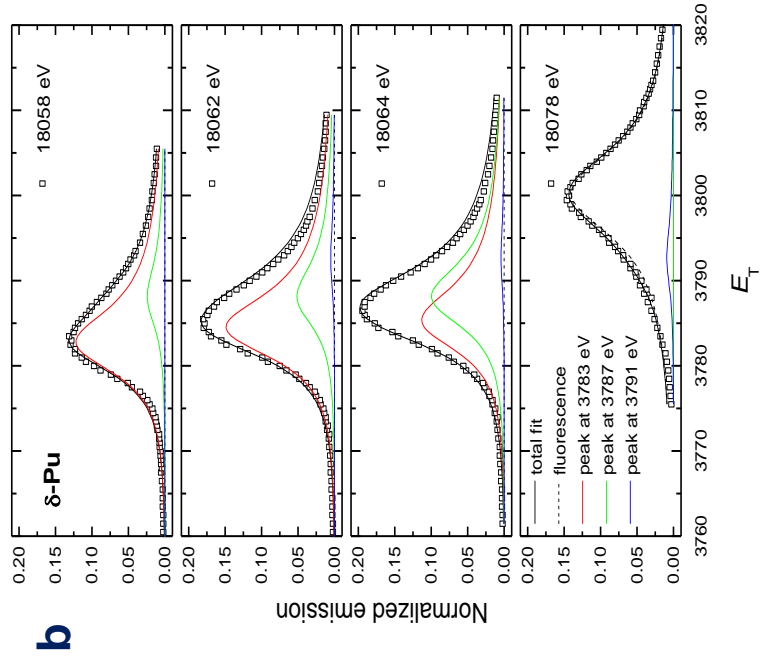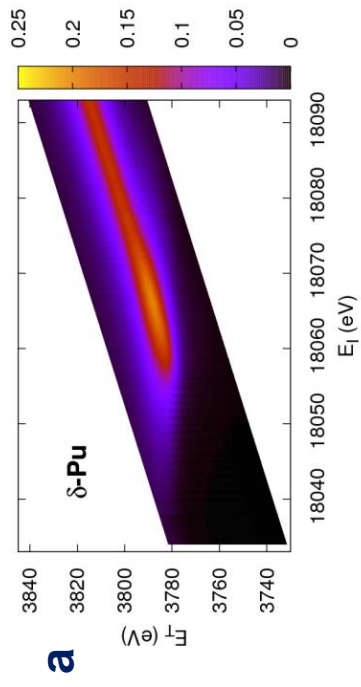

**Figure S3**

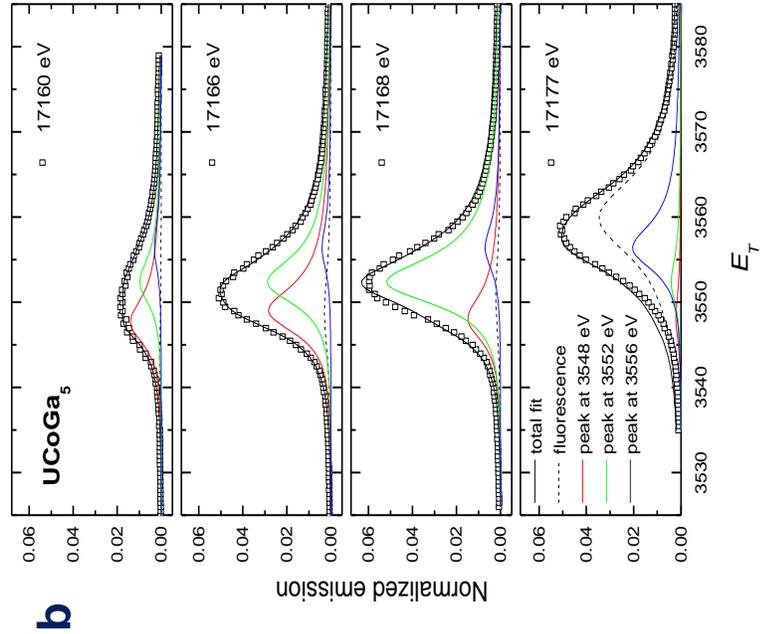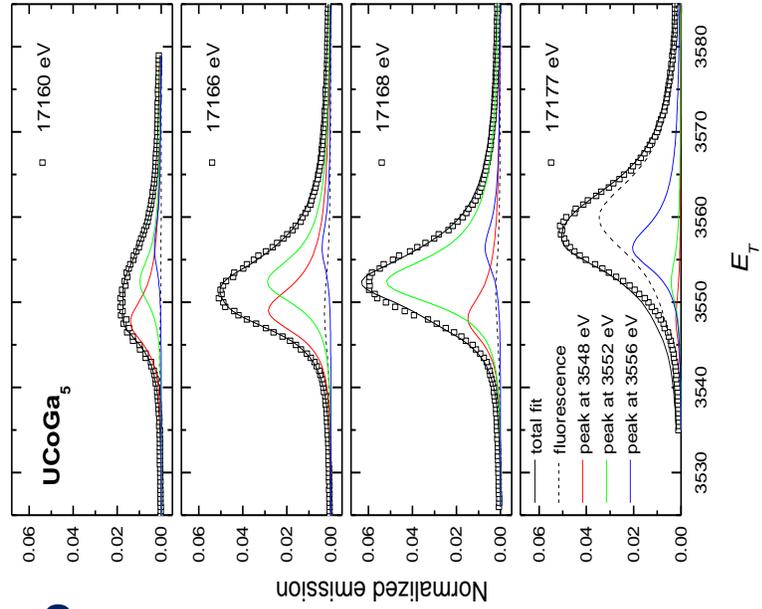

**Figure S4**

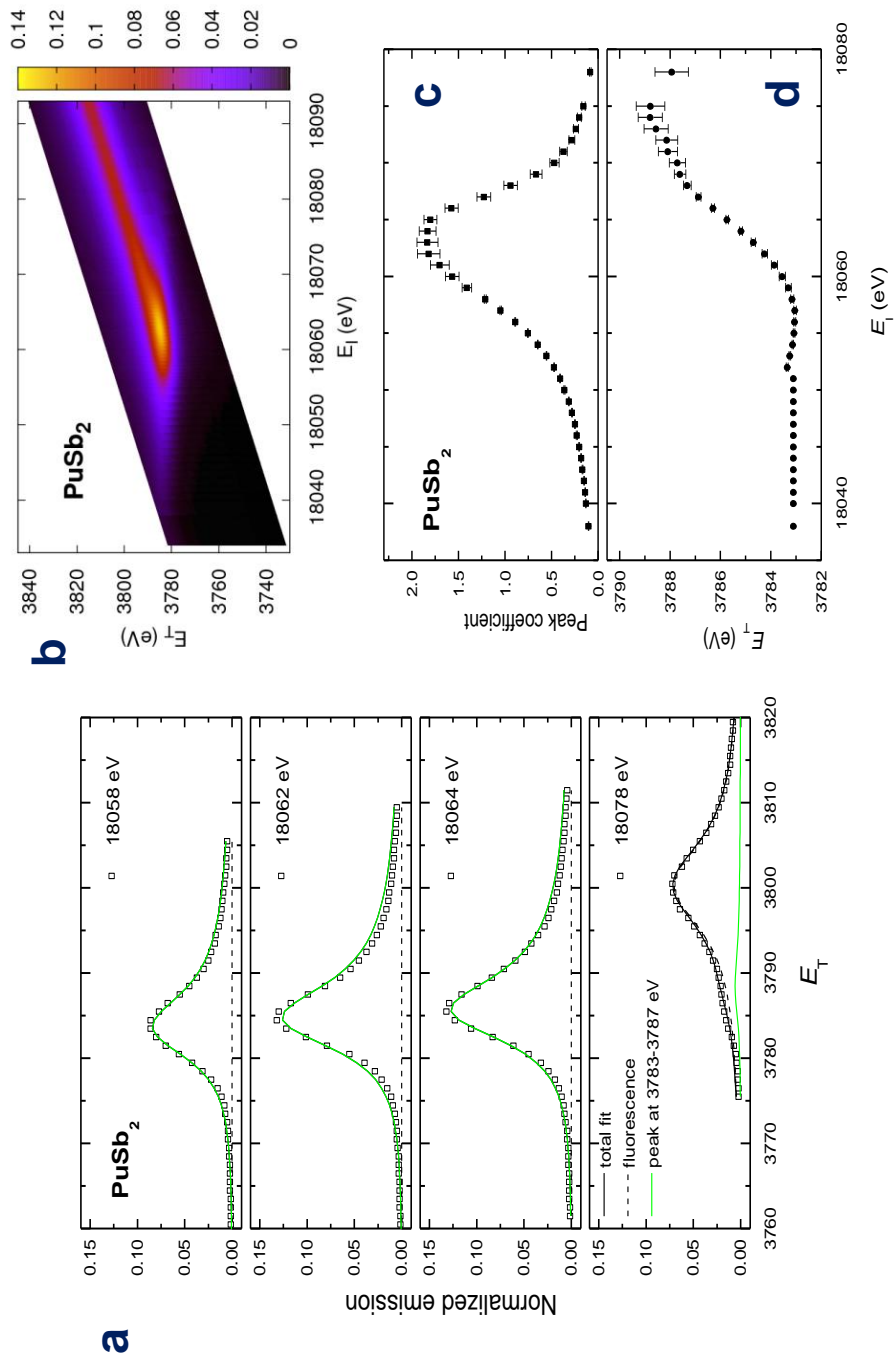

**Figure S5**

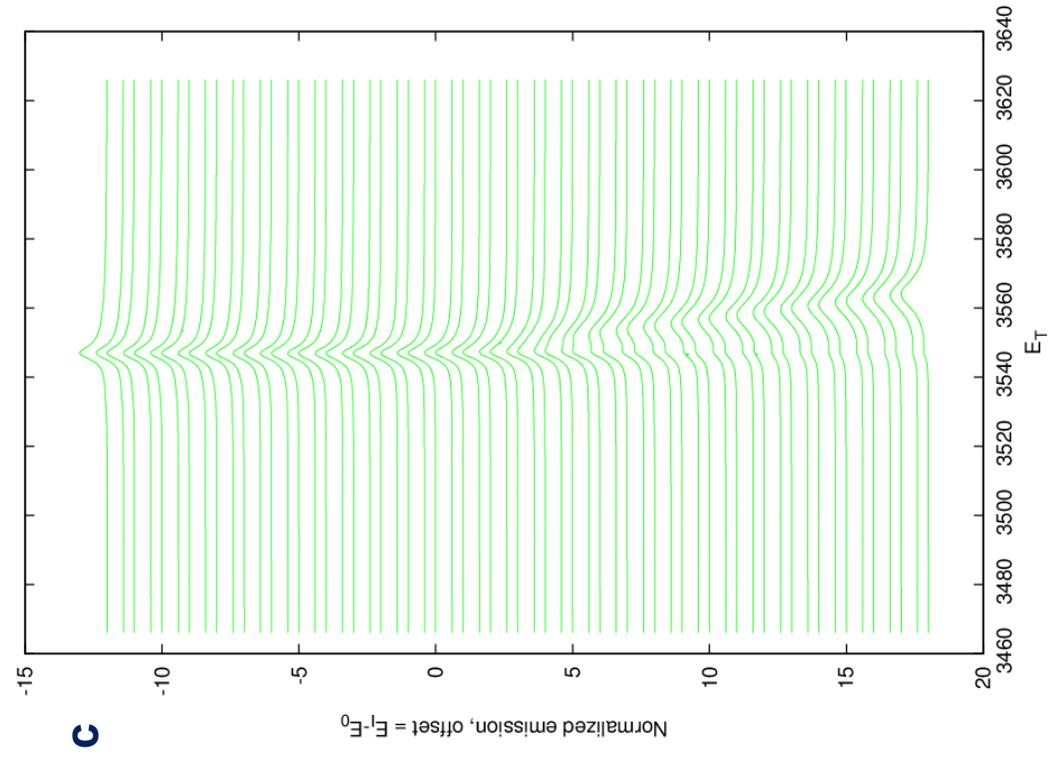
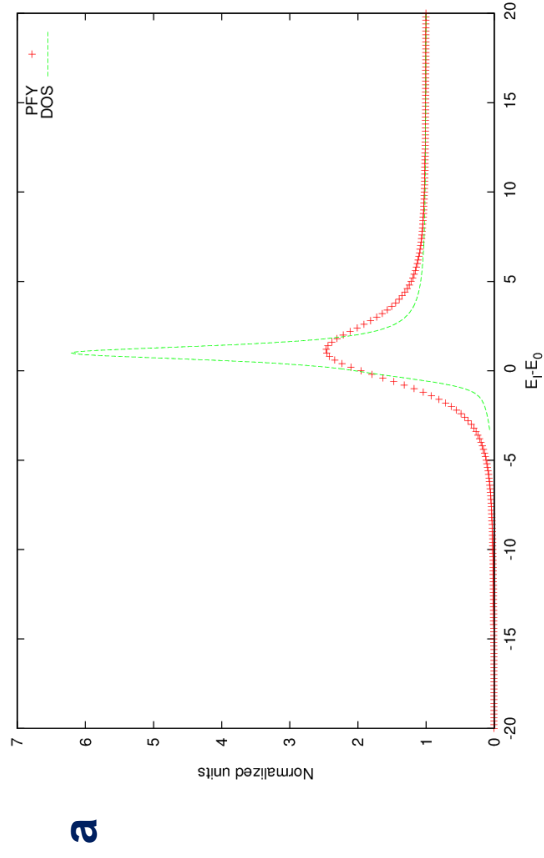
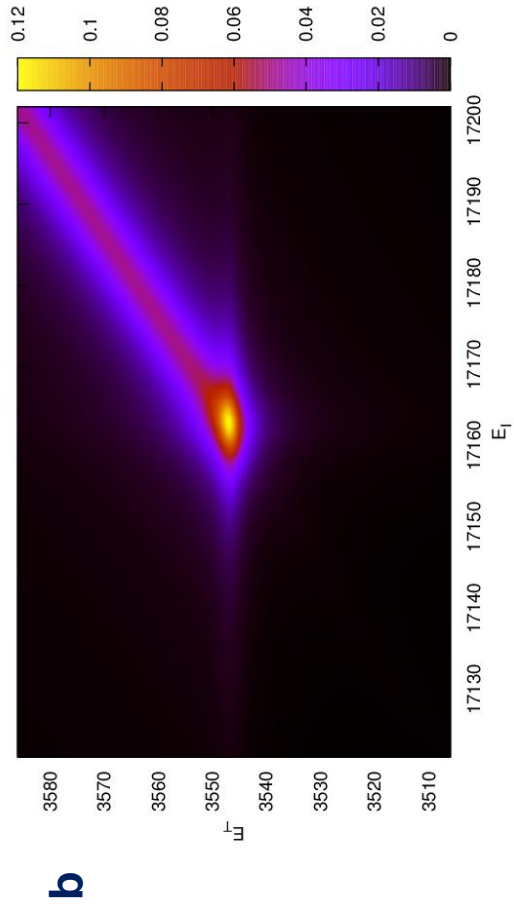

**Figure S6**

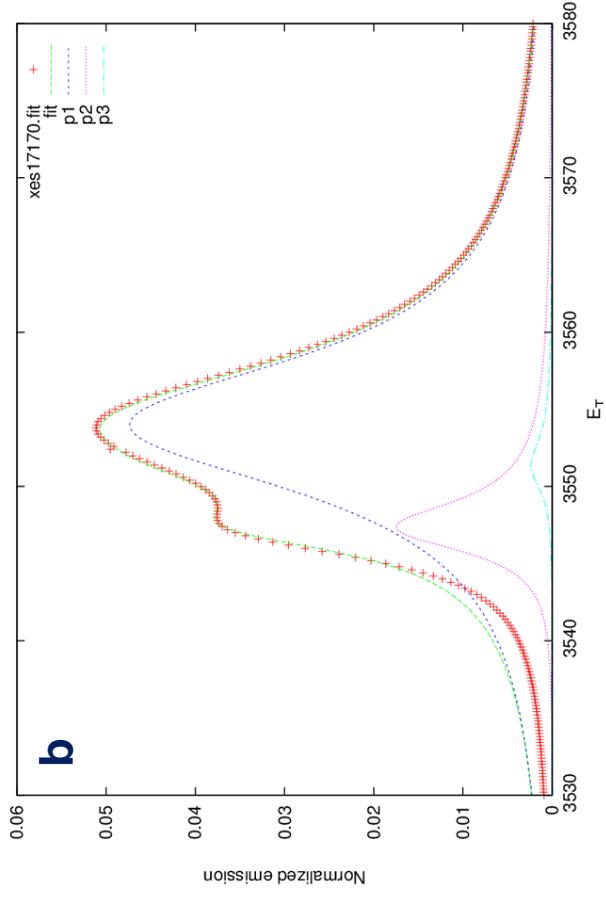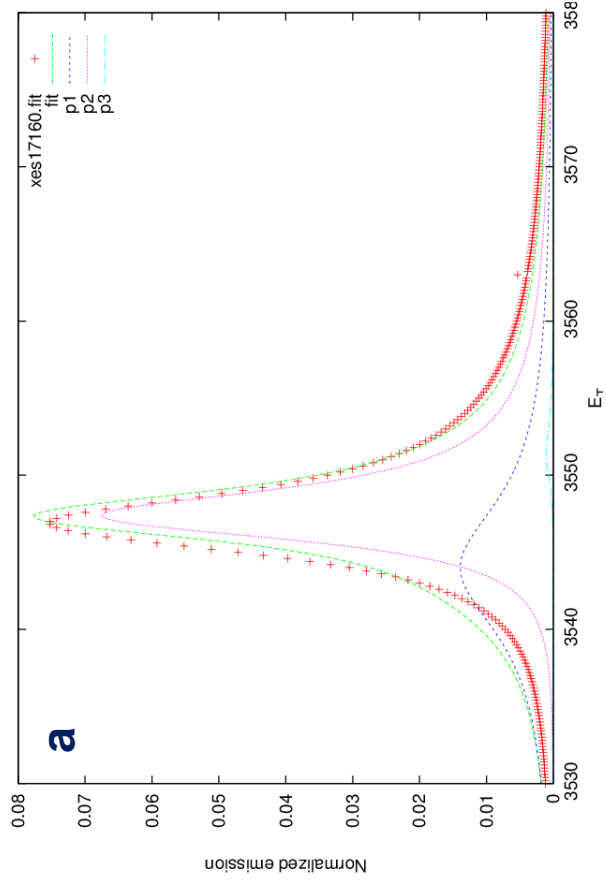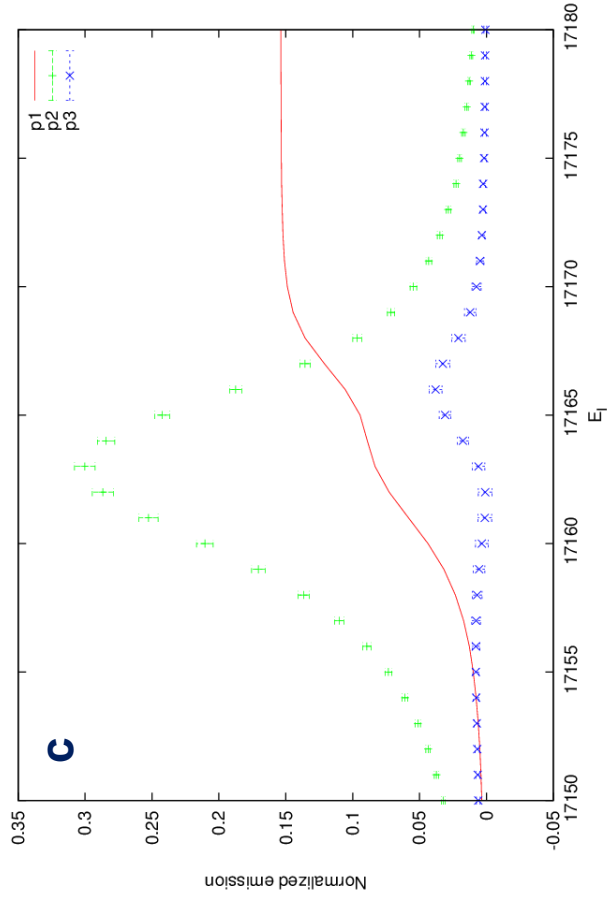

Figure S7